\documentclass[12pt,preprint]{aastex}

\shorttitle{Effects of Rotation on Non-axisymmetric SASI}
\shortauthors{Iwakami et al.}

\begin{document}

\title{Effects of Rotation on Standing Accretion Shock Instability in Nonlinear Phase for Core-Collapse Supernovae}

\author{Wakana Iwakami,\altaffilmark{1}
        Kei Kotake,\altaffilmark{2}
        Naofumi Ohnishi,\altaffilmark{1}
        Shoichi Yamada,\altaffilmark{3,4}
        and
        Keisuke Sawada\altaffilmark{1}}


\email{iwakami@rhd.mech.tohoku.ac.jp}



\altaffiltext{1}{Department of Aerospace Engineering, Tohoku University, 6-6-01 Aramaki-Aza-Aoba, Aoba-ku, Sendai, 980-8579, Japan}
\altaffiltext{2}{Division of Theoretical Astronomy/Center for Computational Astrophysics, National Astronomical Observatory of Japan, 2-21-1, Osawa, Mitaka, Tokyo, 181-8588, Japan}
\altaffiltext{3}{Science \& Engineering, Waseda University, 3-4-1 Okubo, Shinjuku, Tokyo, 169-8555, Japan}
\altaffiltext{4}{Advanced Research Institute for Science and Engineering, Waseda University, 3-4-1 Okubo, Shinjuku, Tokyo, 169-8555, Japan}


\begin{abstract}
We studied the effects of rotation on standing accretion shock instability (SASI) by performing three-dimensional hydrodynamics simulations.
Taking into account a realistic equation of state and neutrino heating/cooling,
we prepared a spherically symmmetric and steady accretion flow 
through a standing shock wave onto a proto-neutron star (PNS). 
When the SASI entered the nonlinear phase,
we imposed uniform rotation on the flow advecting from the outer boundary of the iron core,
whose specific angular momentum was assumed to agree with recent stellar evolution models.
Using spherical harmonics in space and Fourier decompositions in time, we performed mode analysis of the nonspherical deformed shock wave to observe rotational effects on the SASI in the nonlinear phase.
We found that rotation imposed on the axisymmetric SASI did not make any spiral modes and hardly affected sloshing modes,
except for steady $l=2, m=0$ modes.
In contrast, rotation imposed on the non-axisymmetric flow increased the amplitude of spiral modes so that some spiral flows accreting on the PNS were more clearly formed inside the shock wave than without rotation.
The amplitudes of spiral modes increased significantly with rotation in the progressive direction.
\end{abstract}

\keywords{ hydrodynamics --- instability --- neutrinos --- supernovae: general}

\section{Introduction}
Core-collapse supernovae are among the most energetic explosions in the 
universe, catastrophically destroying massive stars.
Because they are relevant to many astrophysical phenomena (e.g., formations of compact stars like neutron stars or black holes, nucleosynthesis, neutrino and gravitational emissions),
their physics has been of wide interest to the astrophysical community.
Regardless of rigorous studies on core-collapse supernovae,
the explosion mechanism is still not completely understood.
Except for the lower mass progenitors,
spherically symmetric supernova simulations have not yet produced explosions \cite[e.g.,][and references therein]{lieben05,sumi}.
As a result of current observations revealing the aspherical nature of the explosion \cite[e.g.,][]{wang02,tanaka,maeda},
multi-dimensional studies and simulations of core-collapse supernovae have explored various mechanisms of asphericity,
such as the roles of convection \cite[e.g.,][]{herant94, burrows95, janka96},
magnetic field and rapid rotation \cite[e.g.,][references therein]{kotake06},
standing accretion shock instability (SASI)
\citep{blondin03,scheck04,blondin06,ohnishi06,ohnishi_2,foglizzo06,ott08},
and $g$-mode oscillations of a proto-neutron star (PNS) \citep{burrows06}.

SASI, the hydrodynamic instability of a standing shock wave,
was originally studied in the context of accreting black holes \citep{foglizzo01,HC}.
The main characteristic of SASI is that lower $l$ modes dominate the flow dynamics,
where $l$ stands for the polar index of the spherical harmonics $Y_{l}^{m}$.
The importance of SASI to the supernova dynamics was first pointed out by 
\citet{blondin03},
who demonstrated that non-radial perturbations added to a standing supernova shock wave grow exponentially with time and lead to $l=1$ or 2 mode deformations in the linear phase, and then
the sloshing motion of the shock wave induced more violent turbulent flows in the nonlinear phase. 
\citet{ohnishi06} indicated that SASI may be the key to the explosion mechanism because SASI can decrease critical neutrino luminosity for the shock revival.
To date, most realistic axisymmetric two-dimensional core-collapse simulations have revealed the appearance of SASI in the initial phase of the explosion \citep{burrows06,marek_janka}.
Moreover, it has been suggested that such a lower mode explosion is favorable for reproducing the synthesized elements of SN1987A \citep{kifo} and for explaining the origins of kick \citep{scheck04}, spins \citep{blondin07a}, and magnetic fields \citep{endeve} of pulsars. 

Reflecting the growing importance of SASI,
the physics behind SASI in supernova cores has recently drawn much attention.
Two types of mechanisms are considered: the advective-acoustic cycle (AAC) and the purely acoustic cycle (PAC).
In the AAC scenario, the entropy and vorticity perturbations advected toward PNS generate a sound wave at the location of the largest velocity gradient of the stationary flow.
The sound wave propagates toward the shock wave and distorts its configuration, depending on the non-radial distribution of the fluctuation pressure.
The deformed shock wave induces further amplification of the entropy and vorticity perturbations \citep{foglizzo07}.
For the PAC scenario, 
the standing pressure wave propagates in the circumferential direction in the region between the spherical accretion shock wave and PNS.
When the postshock pressure is slightly higher than unperturbed pressure, 
it pushes the shock wave outward.
The outward displacement of the shock wave leads to an increase of postshock pressure in the inner region,
while the postshock pressure immediately behind the shock wave decreases.
Thus, the amplitude of the pressure fluctuation increases further \citep{blondin06}.
At this time, which mechanism really works in the supernova cores is still a topic of debate.

For the past few years, the non-axisymmetric features of SASI have been investigated with three-dimensional simulations \citep{blondin07a, iwakami08}.
The modes of SASI are divided into {\it sloshing} modes and {\it spiral} modes.
 Axisymmetric $m = 0$ modes thus far studied in two-dimensional axisymmetric models (symmetric axis is z-axis) and degenerate $|m|= l$ modes (symmetric axes are x-axis, y-axis, and so on) are classified as {\it sloshing} modes,
where $m$ stands for the azimuthal index of the spherical harmonics $Y_{l}^{m}$.
When we impose random perturbation or rotating flow on these axisymmetric flows,
the degeneracy is broken and the rotational modes emerge.
In this situation, the $+m$ modes have different amplitudes than the $-m$ modes.
Such rotating non-axisymmetric $m\neq0$ modes are called {\it spiral} modes.
The sloshing modes can be expressed as degenerate spiral $\pm m$ modes
which have the same amplitudes as each other.
The growth of spiral modes in the linear phase has been examined by simulation with a two-dimensional polar grid of a thin wedge over the entire equatorial plane \citep{blondin07b}.
A deformed shock wave resulting from the growth of the spiral modes produces a spiral accretion flow inside the shock wave, which transfers the angular momentum onto the PNS.
A three-dimensional SASI simulation by \citet{blondin07a} demonstrated that the spiral mode of $m=1$ grew dominantly and generated a strong rotational flow, continuing to develop in the nonlinear phase when random perturbations were imposed on the non-rotating or rotating progenitor.
However, our previous work \citep{iwakami08} revealed that  
the rotational flow did not develop as much for non-rotational models.
An equi-partition was nearly established among different $m$ modes on a time average in the nonlinear phase.
In the flow, high-entropy blobs were produced inside the shock wave.
High-velocity accreting matter ran outside the blobs,
and circulating flows formed inside the blobs.
The high-entropy blobs repeatedly came and went during the nonlinear phase.
In contrast with the outcomes of \citet{blondin07a}, 
our results were similar to those obtained by the Garching group \citep{woosley05}, which revealed that large bubbles of radiation formed inside the shock wave.
Why these results differed so much is not yet understood.

The SASI for the rotating progenitor model has also been studied.
\citet{blondin07a} demonstrated that the rotation of the infalling gas helped to initiate the spiral modes of SASI,
and the rotation axis of the flow was roughly aligned with the spin axis of the progenitor star.
\citet{laming07} and \citet{yamasaki08} analyzed the effect of rotation on SASI.
\citet{laming07} derived an approximate dispersion relation for oscillations of the spherical accretion shock wave.
\citet{yamasaki08} used perturbative analysis with WKB approximation for the cylindrical shock wave. 
Although they devided over the opinion which AAC or PAC worked dominantly, they found that the growth rate of the modes rotating in the same direction as the flow was increased by rotation.
In the present study, we investigated the effect of rotation on SASI in the nonlinear phase.
We introduced a rotation whose axis was along the z-axis into the perturbed flow in the nonlinear phase.
We conducted mode analysis expanding the deformation of the shock wave with spherical harmonics in space and Fourier series in time to distinguish between $+m$ and $-m$ modes.
If the features in the nonlinear phase were similar to the results of linear analysis by \citet{laming07} and \citet{yamasaki08}, it was expected that for clockwise rotation of $\omega_\phi < 0$ the $m < 0$ modes should grow much more than the $m \ge 0$ modes.


The organization of this paper is as follows, we describe the numerical models and the formulations for mode analysis in \S 2, we present the results of computations in \S 3, and we make a summary and discussion of this study in \S 4.

\section{Methods of Computation and Analysis}

\subsection{Numerical Models}
The numerical method was exactly the same as that used in our previous paper \citep{iwakami08}.
Employing the ZEUP-MP code \citep{hayes06} for the hydro solver, 
we solved the dynamics of an accretion flow of matter attracted by the PNS and irradiated by neutrinos emitted from the PNS.
The Shen EOS \citep{shen98} was implemented according to the prescription in \citet{kotake03}. 
To treat the neutrino heating and cooling, 
we used the light bulb approximation \cite[see][for details]{ohnishi06},
in which the neutrino heating is estimated under the assumption that neutrinos are emitted isotropically from the central object with a fixed neutrino flux, and that the matter outside PNS is optically thin.
For simplicity, we considered only the interactions of electron-type neutrinos and anti-neutrinos.
Their temperatures were constantly assumed to be the typical values in the postbounce phase (i.e., $T_{\nu_{\mathrm{e}}} = 4 $~MeV and $T_{\bar{\nu}_{\mathrm{e}}} = 5 $~MeV).
The neutrino luminosity was fixed at $L_\nu = 6.0 \times 10^{52}$~ergs~s$^{-1}$.
The mass of the central objects was assumed to be $M_{\rm in} =1.4 M_\odot$,
and the mass accretion rate was set to be $\dot{M} = 1~M_{\odot}$~s$^{-1}$.

A staggered mesh in the spherical polar coordinates system was used.
The mesh had 300 radial mesh points to cover
$r_{\rm in} \leq r \leq r_{\rm out}$, and 30 polar and 60 azimuthal mesh points to cover the entire solid angle.
Here, $r_{\rm out} = 2000$km was the radius of the outer boundary at which the flow was supersonic,
and $r_{\rm in} \sim 50$km was the radius of the inner boundary located roughly at the neutrino sphere.
This angular resolution was enough to investigate the characteristics of SASI being dominated by lower modes \cite[see][for resolution tests]{iwakami08}.

For the boundary conditions, we imposed the fixed-inflow condition on the outer boundary,
and the free-outflow condition except for the radial velocity $v_r$ on the inner boundary. 
At the points on the inner boundary, $v_r$ was fixed with the value of initial flow in order to obtain a steady flow at the onset of calculation.
At the points on the inner ghost boundary, $v_r$ was determined to satisfy $v_{r,0} = v_{r,1}r^2_1/r^2_0$ $(v_{r,0}>v_{r,1}$), which meant the conservation of mass flux for 1D flow,
where $v_{r,0}$ is the $v_{r}$ at $r_0$ and $v_{r,1}$ is the $v_r$ at $r_1$ $(r_0<r_1)$. 
We confirmed the influence of inner ghost boundary condition to change the relational expression to the fixed-outflow condition ($v_{r,0} = v_{r,1}$) or the fixed-initial-flow condition ($v_{r,0}<v_{r,1}$).
The change of relational expression in the ghost boundary made little difference in the essential features of the flow.

We used the spherically symmetric steady flow as the initial condition \citep{yamasaki05}.
The radial distributions of various variables for the unperturbed flows were given in our previous paper \citep{iwakami08}.
In order to induce non-spherical instability, we added radial velocity perturbation, $\delta v_r(\theta, \phi)$, to
the steady spherically symmetric flow according to the following equation:
\begin{equation}
 v_r(r, \theta, \phi) = v_r^{1D}(r)(1 +\delta v_r(\theta, \phi)),
\end{equation}
where $v_r^{1D}(r)$ is the unperturbed radial velocity.
In this study, we considered two types of perturbation:
(1) an axisymmetric $l=1, m=0$ single-mode perturbation,
\begin{equation}
\delta v_r(\theta, \phi) \propto \sqrt{\frac{3}{4\pi}}\cos\theta,
\end{equation}
and (2) a non-axisymmetric random multimode perturbation,
\begin{equation}
 \delta v_r(\theta, \phi) \propto {\rm rand} \quad  (0 \leq {\rm rand} < 1) ,
\end{equation}
where rand is a pseudo-random number.
These perturbation amplitudes were set to be less than 1\% of the unperturbed velocity.

\subsection{Introduction of Rotation}

In contrast with a spherically symmetric accretion flow,
the method to construct the stationary rotational accretion flow running through the rotationally deformed shock for core-collapse supernovae 
has apparently not been developed thus far.
Hence, to investigate the effects of rotation, we imposed rigid rotation on only the outer boundary.
The imposed rotation is advected toward the shock wave by the accreting flow.
We focused on only the nonlinear phase in this study; therefore, we introduced a rotational flow into the flow field in which SASI had already developed in the nonlinear phase.
We then analyzed the difference between the result with the introduction of the rotation and that without.

We yielded the rotation described as follows on the outer boundary:
\begin{equation}
v_\phi^{2D}(r,\theta) = v_r^{1D} (r) \beta_\phi \sin \theta,
\label{eq:rot}
\end{equation}
where $v_\phi^{2D}(r,\theta)$ denotes the unperturbed $\phi$ component of velocity
and $\beta_\phi$ denotes the rotation parameter.
We examined the flow characteristics for $\beta_\phi=0.005-0.050$, which corresponded to the specific angular momentum ${\it L}\sim 0.2 - 2.3\times10^{16}$~cm$^2$~s$^{-1}$ on the equatorial plane.
Table~\ref{model} lists all the models used in this study.
The $L$ of these models was chosen to be reconciled with the results of the recent presupernova calculations of rotating stars with magnetic fields by \citet{heger05}. 

\begin{figure}
\epsscale{.80}
\plotone{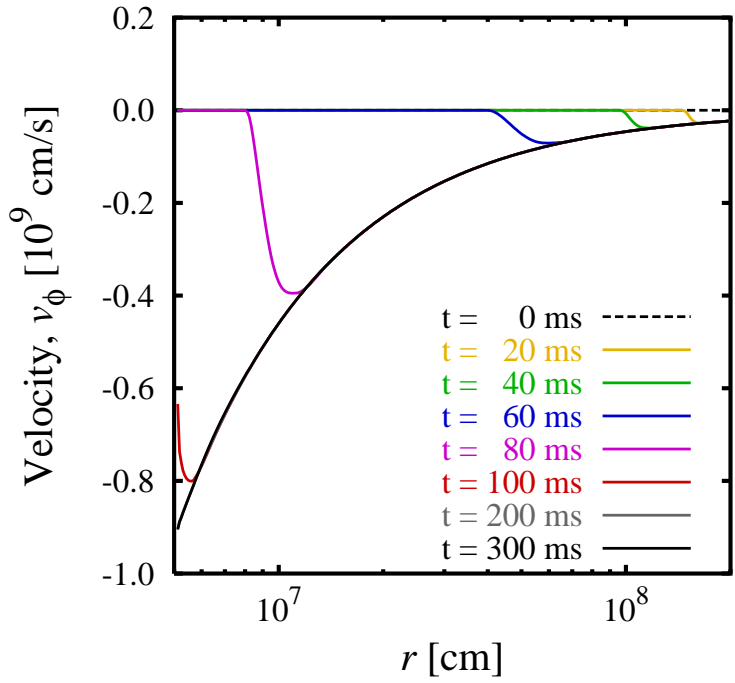}
\caption{
Radial distributions of the $\phi$ component of velocity along the equatorial plane ($\theta = 90^{\circ}, \phi = 0^{\circ}$) for Model A1. 
The initial flow was the steady spherically symmetric one.
The rotation for $\beta_\phi=0.01$ is imposed on the outer boundary at $t=0$ms.
The plots of $t=200$ms and $t=300$ms overlap.
\label{v3}}
\end{figure}

 Figure~\ref{v3} illustrates the evolution of radial distributions of $v_{\phi}$ along the equatorial plane ($\theta = 90^{\circ}, \phi = 0^{\circ}$) for Model A1,
by which one can see how the rotation given to the outer boundary was advected toward the PNS.
At first, $v_\phi$ was zero everywhere 
when the rotational flow for $\beta_\phi=0.01$ was introduced into the outer boundary at $t=0$ms.
The rotational flow with $v_\phi$ was then advected from the outer boundary to PNS with increasing its value by compression for $t=20-100$ms.
At $t=100$ms, the rotational flow reached the surface of PNS and $v_\phi$ hardly changed for $t=150-300$ms, where the plots for $t=200$ms and $t=300$ms overlapped.

\begin{figure}
\epsscale{.60}
\plotone{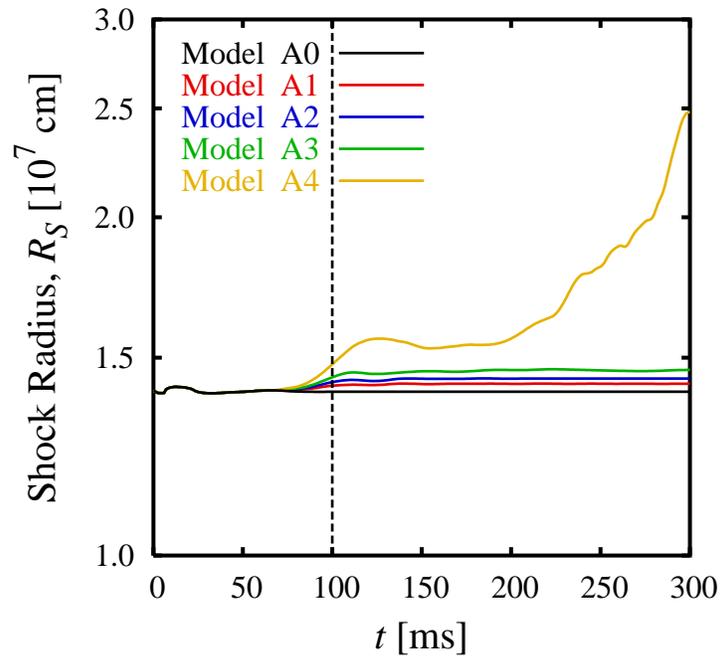}
\caption{
Time evolutions of the average shock radius $R_S$ in the unperturbed flow for Models A0, A1, A2, A3, and A4. The time denoted by a dashed line corresponds to the time of rotation arrival at the inner boundary.
\label{initsr}}
\end{figure}

Figure~\ref{initsr} depicts the time evolutions of the average shock radius $R_S$ for unperturbed flow. 
The average shock radii for Models A0, A1, A2, and A3 were almost constant from 150ms to 300ms.
The average shock radii for the rotating flow of Models A1, A2, and A3 were slightly larger than that for non-rotational Model A0 because of the centrifugal force.
When we applied more rapid rotation like Model A4, the shock wave continued to expand outward.
The models for $\beta_\phi\le0.015$ were the non-exploding models for unperturbed flow, and allowed us to analyze clearly the role of rotational SASI in flow dynamics. Thus, we focused mainly on the models for $\beta_\phi\le0.015$.

\begin{figure}
\epsscale{01.10}
\plottwo{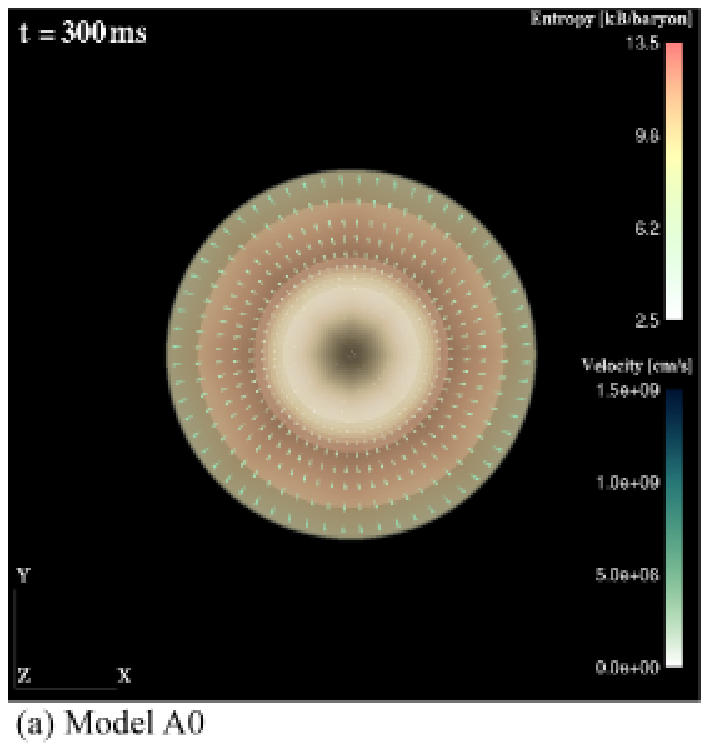}{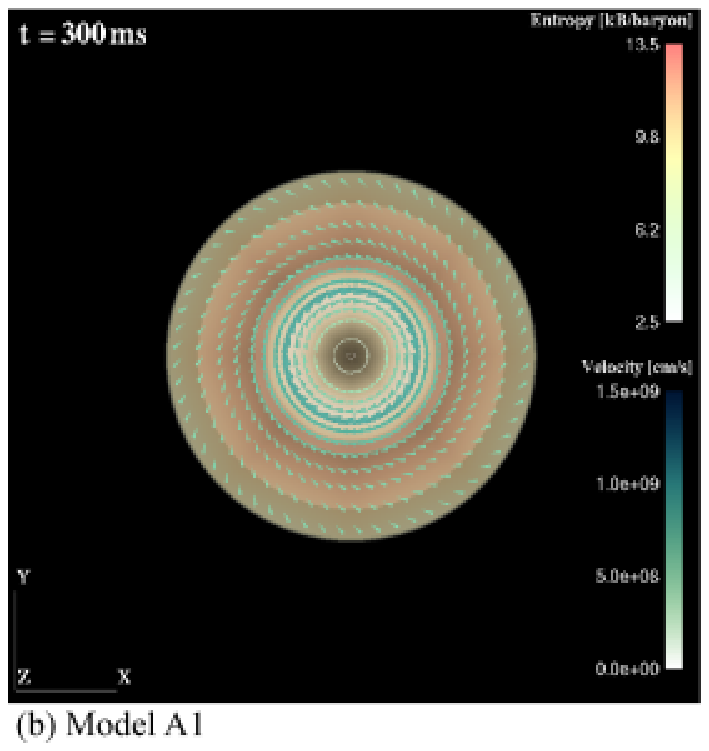}
\caption{
Iso-entropy surfaces and velocity vectors in the equatorial section at $t=300$ms in the unperturbed flow (a) without rotation for Model~A0 and (b) with rotation for Model~A1. The length of each side of a panel corresponds to $5.0\times10^7$cm.
\label{init}}
\end{figure}

Figure~\ref{init} displays the side views of the iso-entropy surfaces with the velocity vectors in the equatorial section.
The entropy is expressed in the eight iso-surfaces shifting from white to the red end of the spectrum with increasing value, and the velocity is denoted by the elongated trigonal pyramids shifting from white to blue.
The hemispheres $(\pi/2\le\theta\le\pi)$ of eight iso-entropy surfaces are superimposed on one another, and the outermost surface almost corresponds to the shock front.
The central region represents the physical quantities on the spherical surface corresponding to the inner boundary.
An unperturbed flow remained spherically symmetric at least until $t\sim 300$ms (Fig.~\ref{init} (a)).
When we introduced the clockwise rotation into the unperturbed flow at $t=0$ms, the rotating flow was advected inward and arrived at the inner boundary at $t\sim 100$ms.
Following that, even though the shock wave slightly oscillated, the flow remained axisymmetric at least until $t\sim300$ms.
This result indicated that no spiral modes were generated numerically (Fig.~\ref{init} (b)).

\subsection{Formulation for Mode Analysis}
This section explains the mode analysis method used in this study.
The deformation of the shock surface could be expanded as a linear combination of the spherical harmonics components $Y^m_{l} (\theta, \phi)$:
\begin{equation}
R_S(\theta, \phi, t) = \sum^{\infty}_{l=0} \sum^{l}_{m=-l} c^m_{l}(t) \, Y^m_{l}(\theta, \phi),
\label{eq:eq8}
\end{equation}
where $Y^m_{l}(\theta, \phi)$ is expressed by the associated Legendre polynomial $P^m_{l}(\cos \theta)$ and
a constant $K^m_{l}$ given as
\begin{equation}
Y^m_{l}(\theta, \phi) = K^m_{l} P^m_{l}(\cos \theta) \, e^{im\phi},
\end{equation}
\begin{equation}
K^m_{l} = \sqrt{\frac{2l+1}{4\pi}\frac{(l-m)!}{(l+m)!}}.
\label{kk}
\end{equation}
The expansion coefficients were obtained by
\begin{equation}
c^m_{l}(t)=\int^{2\pi}_0 \! \! \! \! d\phi  \! \int^{\pi}_0 \! \! d\theta \, \sin \theta \, R_S(\theta, \phi, t) \, Y^{m*}_{l} (\theta, \phi),
\label{clm}
\end{equation}
where the superscript * denotes complex conjugate.
Instead of the expansion coefficients $c^m_{l}(t)$, 
we used normalized amplitudes $c^m_{l}(t)/c^0_{0}(t)$ for analysis.

Moreover, normalized amplitudes $c^m_l(t)/c^0_0(t)$ could be expanded to the Fourier series as follows:
\begin{equation}
c^m_{l}(t)/c^0_{0}(t)=\int^{\infty}_{-\infty} \! \! \! \! d\omega \, \hat{c}^m_l(\omega) \, e^{-i\omega t},
\end{equation}
where $\omega$ is a real number denoting an oscillation frequency.
Thus, we could rewrite (\ref{eq:eq8}) as
\begin{equation}
R_S(\theta, \phi, t)/c^0_0(t) = \sum^{\infty}_{l=0} \sum^{l}_{m=-l} \int^{\infty}_{-\infty} \! \! \! \! d\omega\, \, \hat{c}^m_{l}(\omega) \, K^m_{l}P^m_{l}(\cos \theta) \, e^{-i(\omega t - m\phi)},
\label{eq:eq13}
\end{equation}
Therefore the Fourier expansion coefficients $\hat{c}^m_l(\omega)$ were calculated as
\begin{equation}
\hat{c}^m_{l}(\omega)=\frac{1}{t_e-t_s}\int^{t_e}_{t_s} \! \! \! \! \! dt \int^{2\pi}_0 \! \! \! \! d\phi  \! \int^{\pi}_0 \! \! d\theta \, \sin \theta \, \left[R_S(\theta, \phi, t)/c^0_0(t)\right] \, K^{m}_{l}P^{m}_{l} (\cos \theta) \, e^{i(\omega t - m \phi)},
\label{eq:cwlm}
\end{equation}
where $t_s$ is the starting time of sampling, and $t_e$ is the ending time.
The Fourier expansion of $c_l^m(t)/c_0^0(t)$ allowed $\hat{c}^m_l(\omega)$ to distinguish between $+m$ and $-m$ modes.

\section{Results}

\begin{figure}
\epsscale{0.60}
\plotone{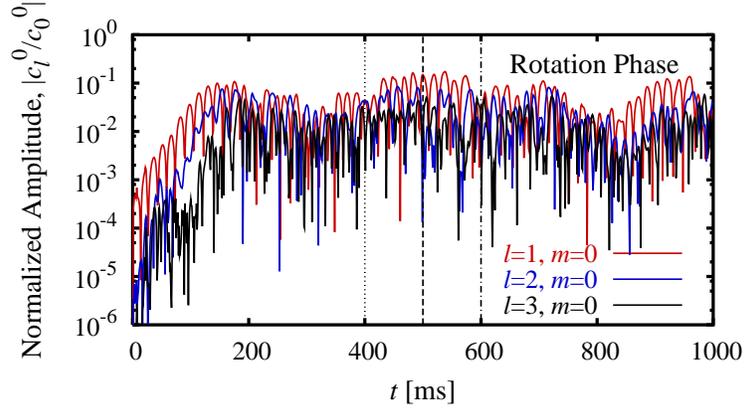}
\vspace{10mm}
\plotone{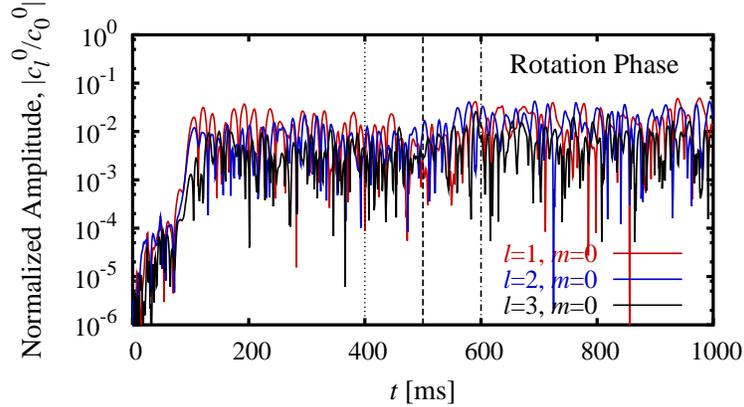}
\vspace{10mm}
\caption{
Time evolutions of the normalized amplitudes $|c^m_l(t)/c^0_0(t)|$. We imposed (a) axisymmetric $l=1,m=0$ perturbation for 2D Model B1 and (b) non-axisymmetric random perturbation for 3D Model~C1 on the spherical symmetric flow at $t=0$ms.
The dotted line indicates the starting time to impose rotation on the bouter boundary,
the dashed line indicates the time of rotation arrival at the inner boundary,
and the dot-dashed line indicates the onset of the rotation phase.
\label{sph}}
\end{figure}

In this study, we introduced the rotation described above into axisymmetric (2D) and non-axisymmetric (3D) flows in the nonlinear phase at $t=400$ms to observe the behavior of the flow in only the nonlinear phase.
We made both 2D and 3D flows in the nonlinear phase with three-dimensional simulation as follows.
Figure~\ref{sph} presents the plots of the normalized amplitudes $|c^m_l(t)/c^0_0(t)|$ as a function of time for 2D Model B1 and 3D Model C1.
In 2D Model~B1 (Fig.~\ref{sph} (a)),
the axisymmetric $l=1, m=0$ perturbation was imposed on the initial flow.
As we described in our previous paper \citep{iwakami08},
the amplitude of the $l=1, m=0$ mode grow exponentially in the linear phase,
and the growth of amplitudes of all modes is saturated in the nonlinear phase.
In 3D Model~C1 (Fig.~\ref{sph} (b)), the non-axisymmetric random perturbation was imposed on the initial flow.
As we mentioned in the same paper, the amplitudes of all the modes grow from the beginning, and then the flow enter the nonlinear phase.
For both 2D and 3D flows, we imparted the rotation on the outer boundary at $t=400$ms (dotted line).
At that time, the flow was completely in the nonlinear phase. 
The rotational flow was carried by an accretion flow and arrived at the inner boundary around $t=500$ms (dashed line).
Following that, the flow inside the shock wave changed, reflecting the infalling rotational flow.
We defined the period from 600ms (dot-dashed line) to 1000ms as the rotation phase.
We integrated the shock radii $R_S(\theta, \phi)$ from 600ms to 1000ms in Eq.~(\ref{eq:cwlm}) for the Fourier expansion.

\subsection{Axisymmetric Rotational Model}

First, we will discuss the results of axisymmetric flows.
All the simulations were carried out in three-dimensions even for the axisymmetric flow.
We dare to term the axisymmetric flow as the 2D flow not to confuse axisymmetric flow with axisymmetric mode.
When we introduced the $l=1,m=0$ perturbation into the non-rotational flow, 
only $m=0$ modes grew, and the flow retained symmetry with respect to the z-axis even in the nonlinear phase.
In other words, the sloshing modes grew, but the spiral modes did not \citep{iwakami08}.
Confirming that only sloshing modes appeared even for the rotational models,
hereafter, we focused on the effect of rotation on the sloshing modes in the nonlinear phase.
Furthermore, we examined that the results of the new mode analysis presented in Section 2.3 to expand the geometry of the deformed shock wave with spherical harmonics in space and Fourier transform in time,
which were consistent with those of the previous mode analysis to expand with only spherical harmonics in space at the given time.

\begin{figure}
\epsscale{0.60}
\plotone{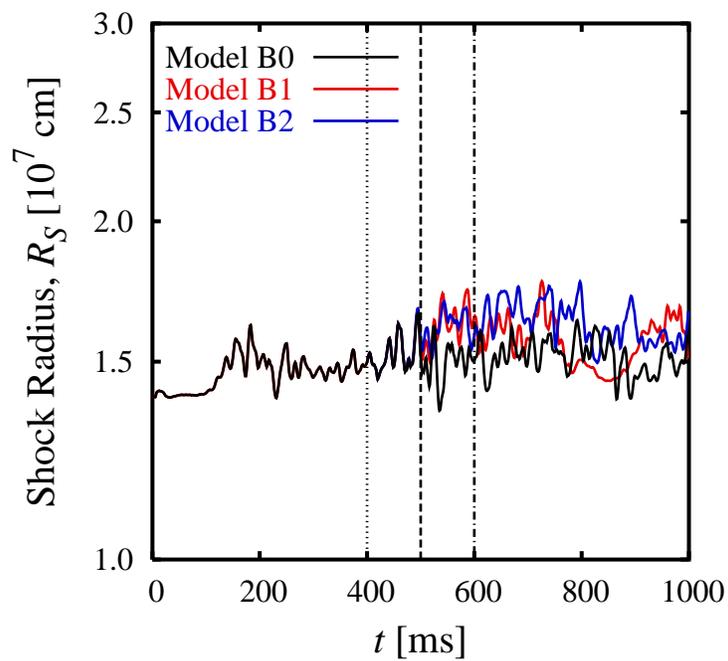}
\caption{
Time evolutions of the average shock radius $R_S$ without rotation for Model~B0 and with rotation for Models B1 and B2. 
The times denoted by the dotted line, dashed line, and dotted-dashed line are the same as in Fig.~\ref{sph}.
\label{srlm10}}
\end{figure}

Figure~\ref{srlm10} depicts the time evolutions of the average shock radius $R_S$ for 2D Models~B0, B1, and B2.
The average shock radius initially increased with the growth of SASI, and then it remained roughly constant until 500ms with little fluctuation.
After 500ms, rotation affected the flow dynamics for Models B1 and B2.
The magnitudes of average shock radii were not as different between non-rotational Model B0 and rotational Models B1 and B2.
We observed that the shock wave burst suddenly if we imposed more rapid rotation on the flow.
In this paper, we focused on only the non-exploding Models B1 and B2.

\begin{figure}
\epsscale{1.10}
\plottwo{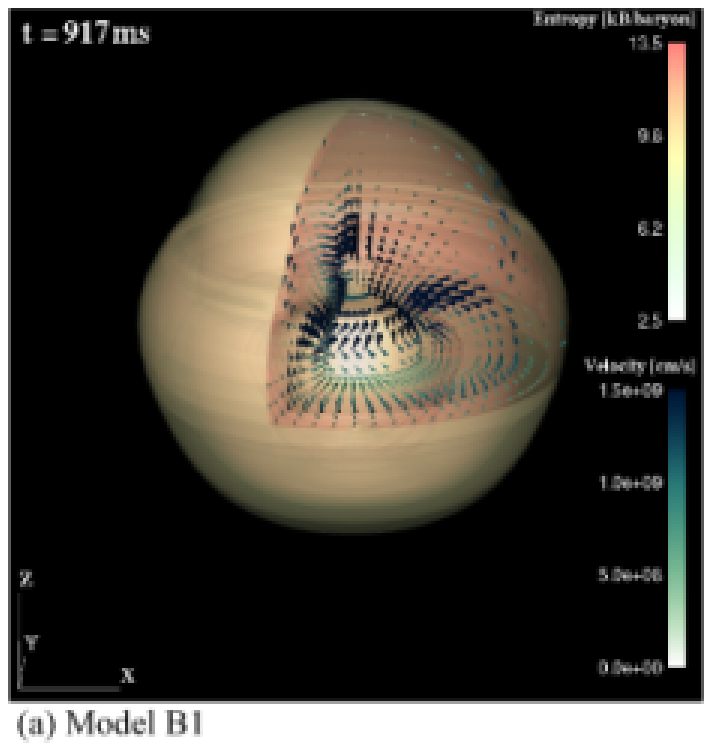}{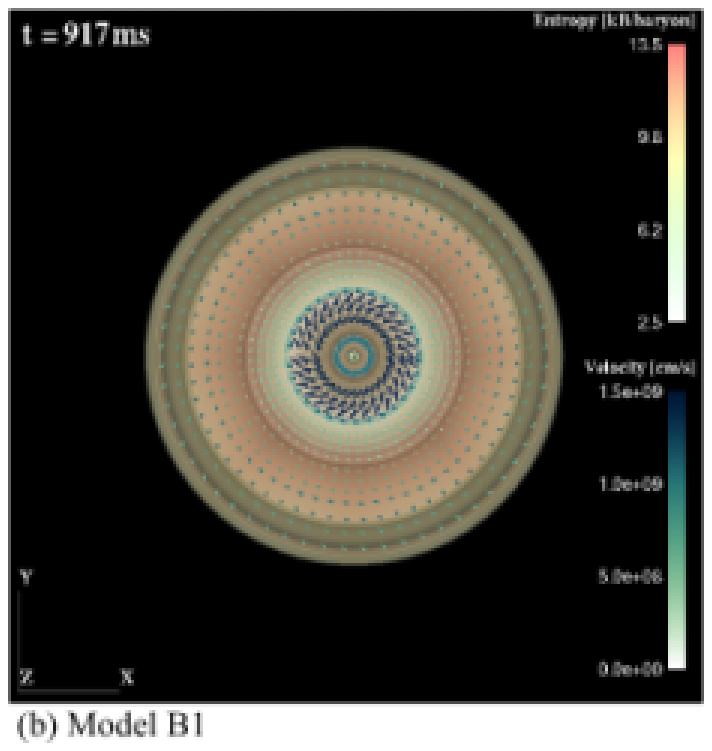}
\caption{
Partial cutaway view of the iso-entropy surfaces and the velocity vectors on the cutting plane at $t=917$ms with rotation for Model~B1.
(a) The object having three cutting planes ($\theta = 90^{\circ}, \phi = 0^{\circ}, 250^{\circ}$) viewed from the side of $-y$ direction and (b) its equatorial section of $\theta = 90^{\circ}$ viewed from $z$ direction.  }
\label{lm10}
\end{figure}

The flow fields for rotational 2D Model~B1 are presented in Fig.~\ref{lm10}.
The velocity vectors are superimposed on the cutting plane of the partial cutaway view of the iso-entropy surfaces.
As with the non-rotation, the flow field retained symmetry with respect to z-axis even in the flow rotating globally (Fig.~\ref{lm10} (a)).
No spiral flows formed, so the flow just rotated clockwise (Fig.~\ref{lm10} (b)).
Thus, it was clear that the sloshing modes grew and that no spiral modes originated with numerical errors emerged.

\begin{figure}
\epsscale{0.60}
\plotone{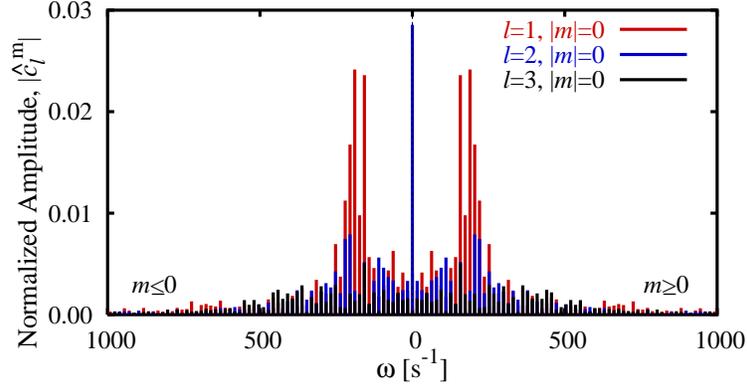}
\vspace{10mm}
\plotone{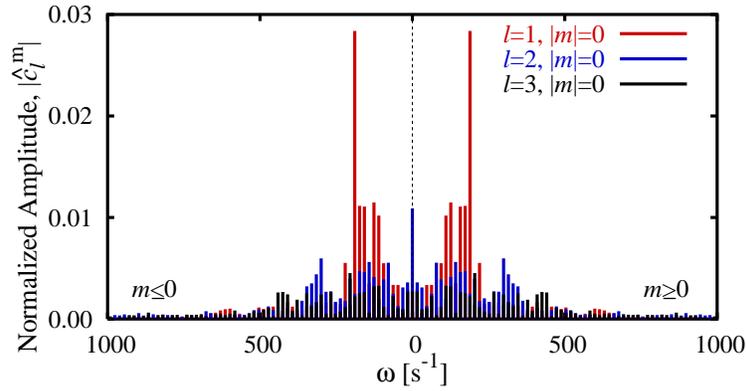}
\vspace{10mm}
\plotone{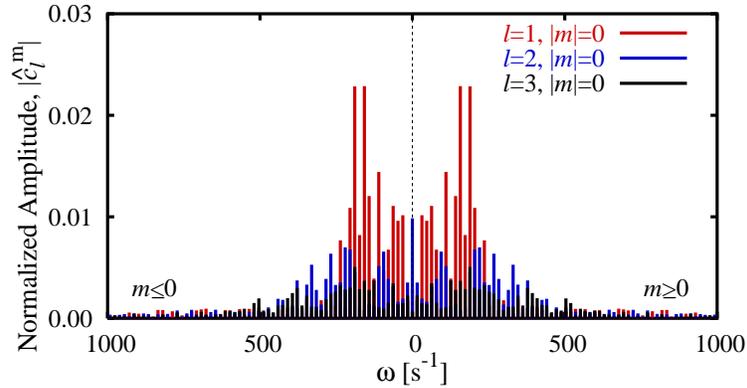}
\vspace{10mm}
\caption{
Fourier-transformed normalized amplitude $|\hat{c}^m_l(\omega)|$ of sloshing $m=0$ modes without rotation for (a) Model~B0 and with rotation for (b) Model~B1 and (c) Model~B2. Normalized amplitudes were estimated by integrating the shock deformation during the rotation phase.
\label{sphflm10}}
\end{figure}

Figure~\ref{sphflm10} presents the Fourier-transformed normalized amplitudes $|\hat{c}^m_l(\omega)|$ of sloshing $m=0$ modes for 2D Models B0, B1, and B2.
The $m\ge0$ modes are plotted on the right side of the graph, and the $m\le0$ modes are plotted on the left side.
In this figure, the plots on both sides are same due to $m=0$, in other words, the left side is just a copy of the right side.
The results of the new mode analysis indicated that the lower modes tended to be dominant, consistent with the feature obtained by the previous mode analysis presented in Fig.~\ref{sph} (a).
The magnitude of the maximum amplitudes of $l=1$ mode and its oscillation frequency $\omega$ were almost the same between non-rotational Model~B0 and rotational Models~B1 and B2.
These results indicated that the rotation hardly affected the characteristics of the sloshing modes in SASI.
However, the steady $l=2, m=0$ mode was much smaller for rotational Models~B1 and B2 than for non-rotational Model~B0.
This may come from the fact that centrifugal force acted more strongly on the matter around the equatorial plane than near the poles.
The vertically long sllipsoidal shock wave oscillating above and below for Model~B0 might be transformed into the more spherical configuration of Models~B1 and B2 by rotation.

\subsection{Non-Axisymmetric Rotational Model}

Next, we considered the results of non-axisymmetric flows.
Here, we call the non-axisymmetric flow as the 3D flow.
When we imposed a random perturbation on the non-rotational flow,
both axisymmetric $m=0$ modes and non-axisymmetric $m\neq0$ modes grew \citep{iwakami08}.
In this situation, both sloshing modes and spiral modes could be generated.
However, our previous mode analysis was not able to distinguish between sloshing modes and spiral modes for non-axisymmetric $m\neq0$ modes,
because the spectra were obtained from the instantaneous deformation of the shock wave.
So we used a new mode analysis in an attempt to confirm whether spiral modes were actually generated.
Furthermore, we investigated the effect of rotation on sloshing and spiral modes in the nonlinear phase.

\begin{figure}
\epsscale{0.60}
\plotone{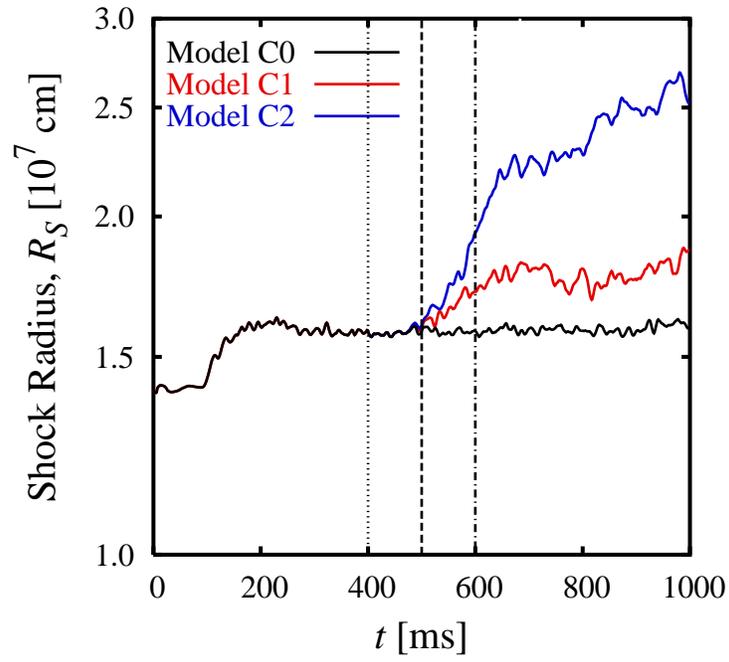}
\caption{
Time evolutions of the average shock radius $R_S$ without rotation for Model~C0 and with rotation for Models C1 and C2. 
\label{srrand}}
\end{figure}

Figure~\ref{srrand} plots the average shock radius $R_S$ as a function of time for 3D Models C0, C1, and C2.
As with 2D Model B0, the average shock radius increased with growing SASI from 100ms,
and then it remained nearly constant until 500ms.
At 500ms, the flow started to come under the influence of rotational flow falling to PNS for Models C1 and C2.
Unlike the 2D models, the faster the rotation, the larger the average shock radius tended to be.
When we imposed more rapid rotation to the flow than in Model C2, the expansion of the shock wave continuously proceeded to the outer boundary.
In this study, we therefore focused on only the non-exploding models of Models C1 and C2.

\begin{figure}
\epsscale{1.10}
\plottwo{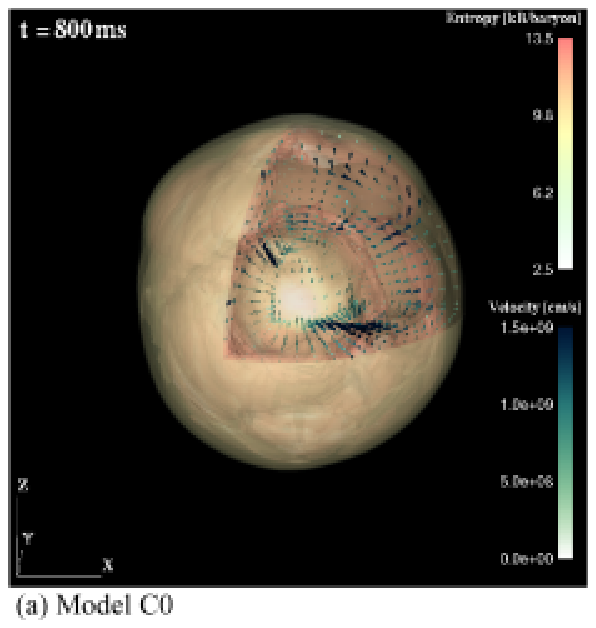}{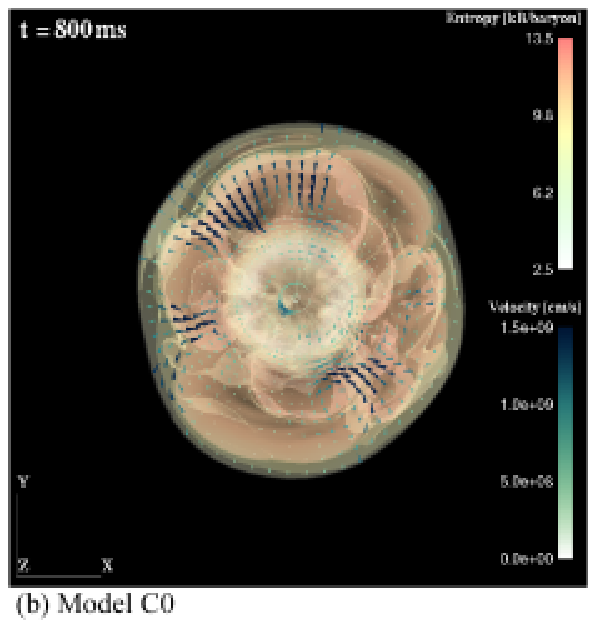}
\plottwo{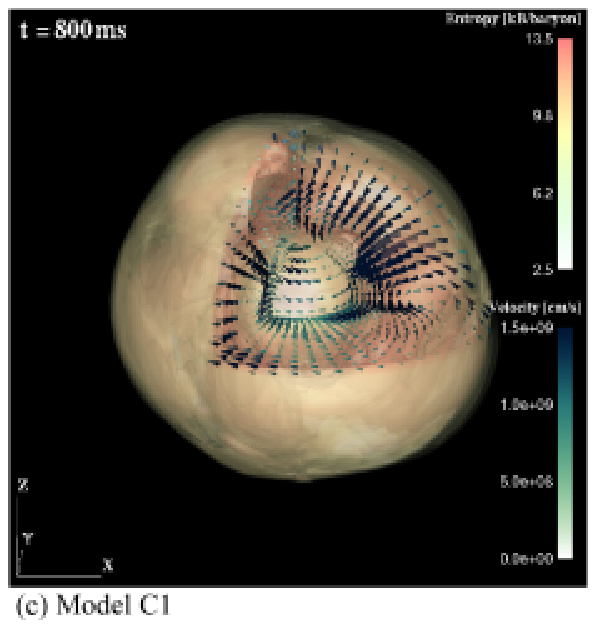}{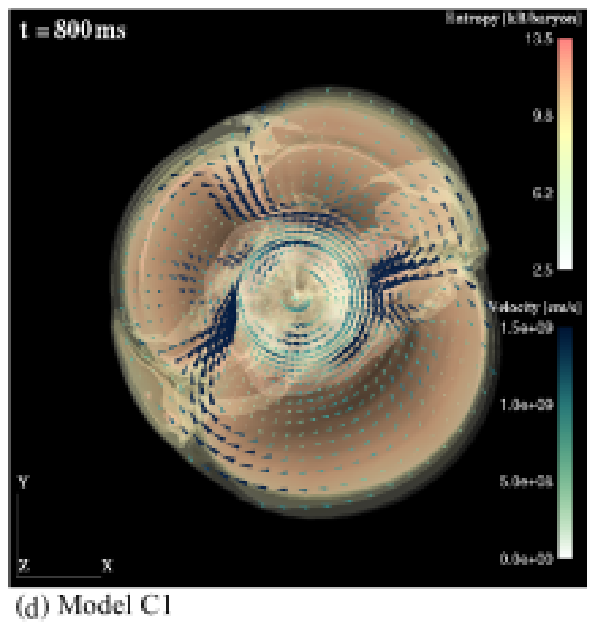}
\caption{
Partial cutaway view of the iso-entropy surfaces and the velocity vectors on the cutting plane at $t=800$ms (a), (b) without rotation for Models C0 and (c), (d) with rotation for Model C1.
One can see (a), (c) the object having three cutting planes from the $-y$ direction and (b), (d) its equatorial section from $z$ direction.}
\label{rand}
\end{figure}

Figure~\ref{rand} illustrates the flow fields for non-rotational 3D Model C0 and for rotational 3D Model C1. 
Symmetry with respect to the z-axis was broken for both models (Figs.~\ref{rand}(a), (c)).
Many large high-entropy blobs were observed inside the shock wave.
The blobs with circulating flow arised and expanded with growing SASI, and pushed away the shock wave outward.
Matter infalling from a triple point on the shock wave ran through the interstices of the blobs with high velocity, and the stream accreted on the PNS in an arc (Figs.~\ref{rand}~(b), (d)).
A triple point is the connection point (or segment) of two shock waves propagating from both sides.
When rotation was not imposed, no specific rotation axis was observed in the flow.
On the other hand, as a result of introducing the rotation whose axis corresponded to the z-axis, the flow inside the shock wave rotated globally around the z-axis, and accreting spiral flows had higher velocity than that for non-rotational models because of $\phi$ component of velocity.
After the accreting flows turned in the rotational direction near the PNS, the flows went up inside the high-entropy blobs with higher velocity than that for non-rotational models (Figs.~\ref{rand}~(d)).
This is favorable for the shock expansion.
Moreover, after the upward flow with high velocity turned in the rotational direction along the shock front,
the centrifugal forces acted on the flow so that the shock wave could be pushed away furthermore (Figs.~\ref{rand}~(d)).
These flow behaviors might be why the average shock radius for rotational 3D models tended to be larger than that for non-rotational ones (Fig.~\ref{srrand}),
while the average shock radii were not different so much between rotational and non-rotational 2D models (Fig.~\ref{srlm10}).
The flows accreting from the triple points with high velocity were developed in the $\theta$ direction for 2D models (Fig.~\ref{lm10}~(a)).
Thus, the flows running in the rotational direction for 2D rotational models do not have so much high velocity as that for 3D ones (Fig.~\ref{lm10}~(b)).
Centrifugal forces, therefore, less acted on the 2D flow than the 3D one so that the shock wave might hardly expanded for 2D models.

\begin{figure}
\epsscale{0.60}
\plotone{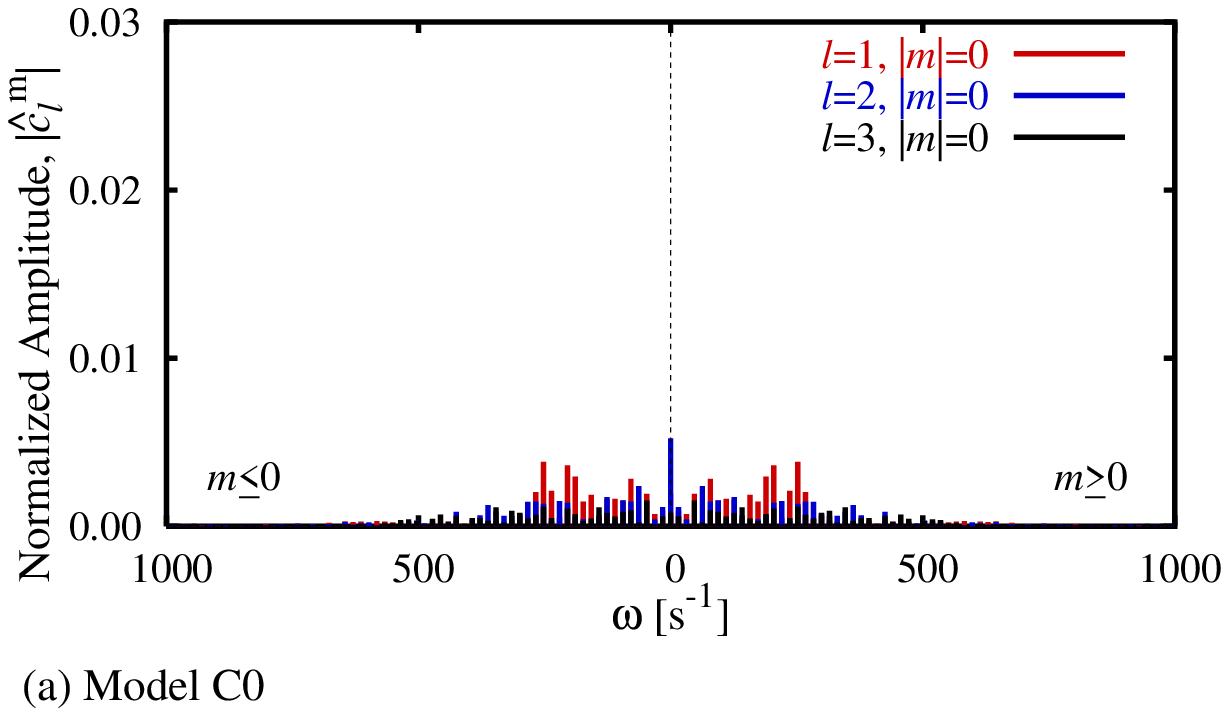}
\vspace{10mm}
\plotone{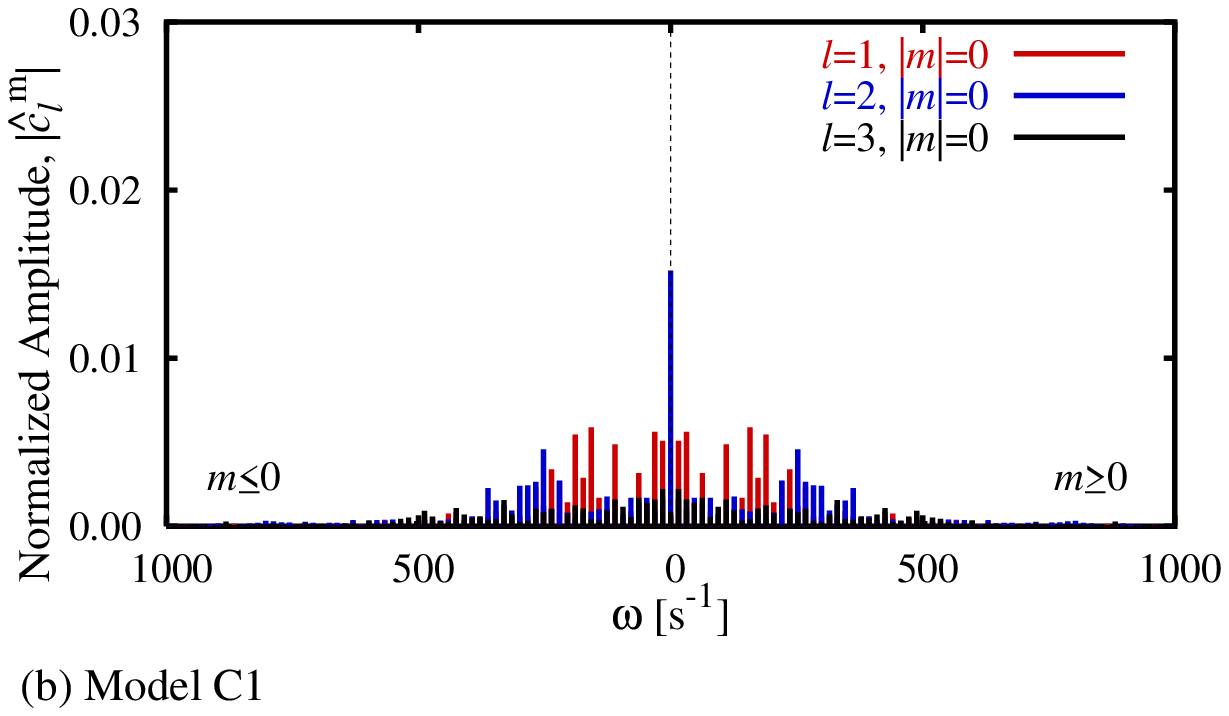}
\vspace{10mm}
\plotone{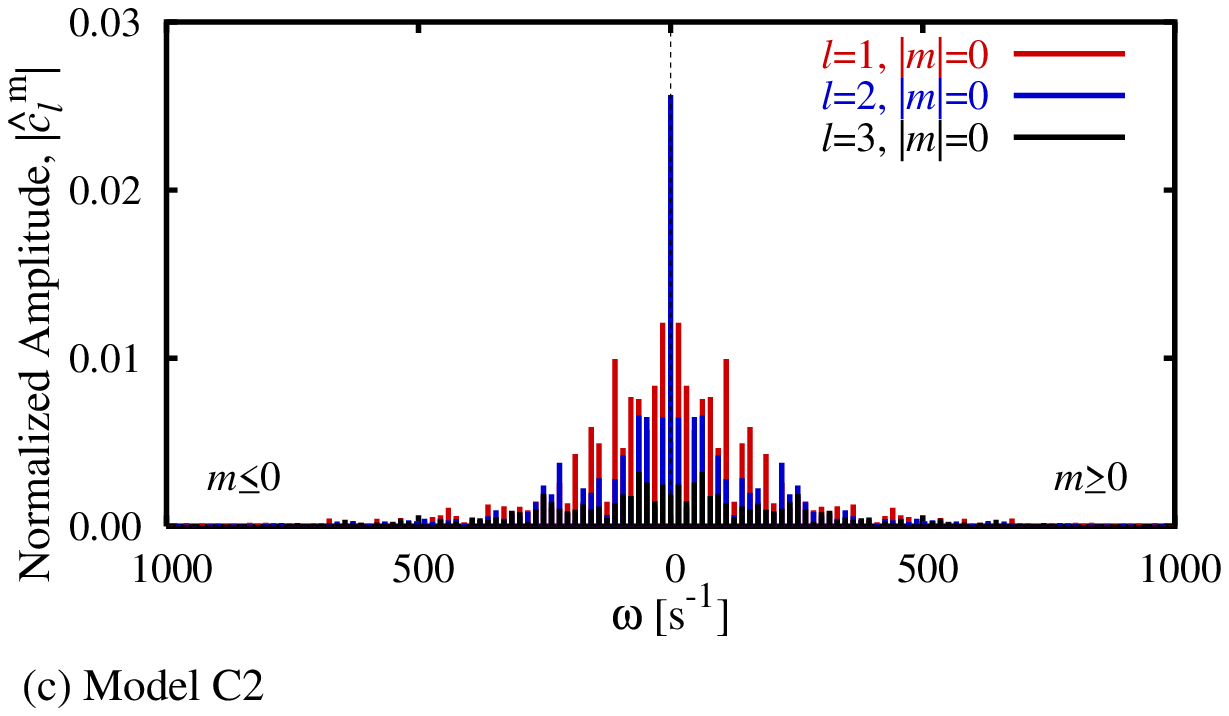}
\vspace{10mm}
\caption{
Fourier-transformed normalized amplitude $|\hat{c}^m_l (\omega)|$ of sloshing $m=0$ modes without rotation for (a) Model~C0 and with rotation for (b) Model~C1 and (c) Model~C2.
\label{sphfrandl0}}
\end{figure}

Figure~\ref{sphfrandl0} depicts the Fourier-transformed normalized amplitudes $|\hat{c}^m_l(\omega)|$ of sloshing $m=0$ modes.
Figure~\ref{sphfrandl0}~(a) presents the results of non-rotational 3D Model~C0.
Lower modes were dominant.
The amplitudes for 3D Model~C0 tended to be smaller than those for 2D Model~B0 (Figs.~\ref{sphflm10}~(a) and Fig.~\ref{sphfrandl0}~(a)).
These results also did not conflict with those of previous mode analyses (Figs.~\ref{sph}~(a), (b)).
Figures~\ref{sphfrandl0}~(b) and (c) present the results of rotational 3D Models C1 and C2.
The steady $l=2, m=0$ sloshing modes grew significantly.
The spherical shock wave for non-rotational models might change into a horizontally long ellipsoidal one because of the centrifugal forces.
The amplitudes of the other sloshing modes also became somewhat large in response to the increase of rotational velocity,
perhaps due to the nonlinear effects of coupling the $m=0$ and $m\ne 0$ modes.

\begin{figure}
\epsscale{0.60}
\plotone{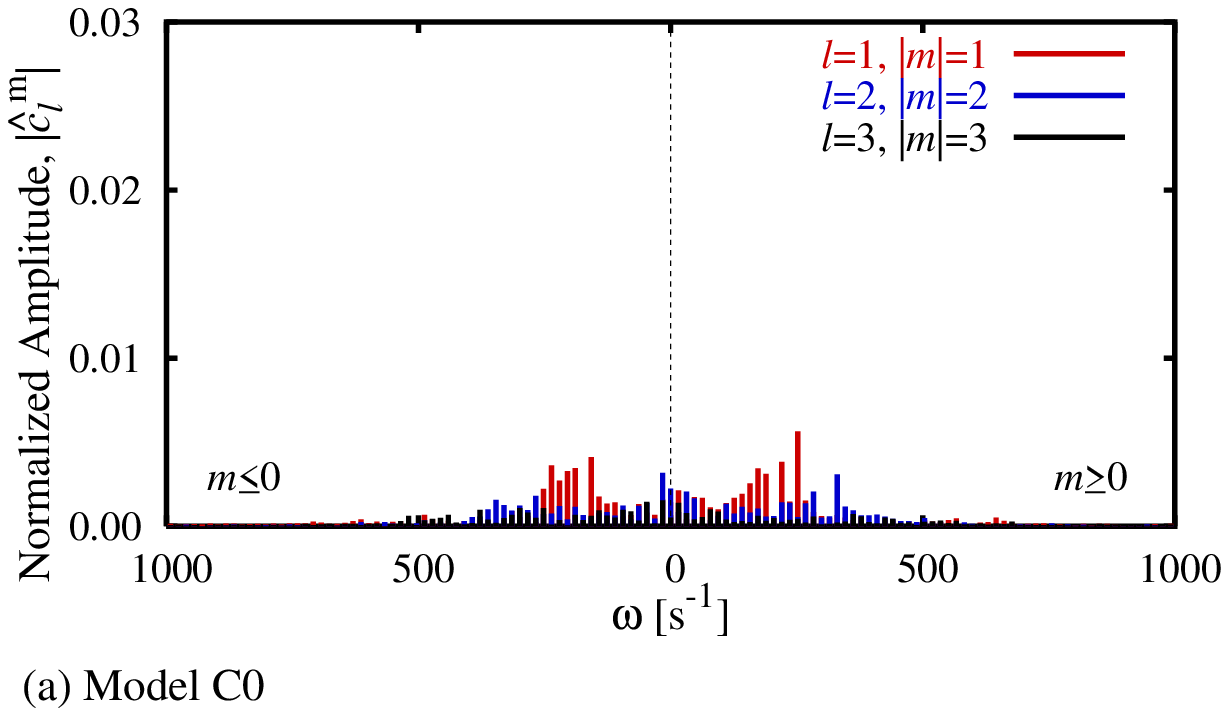}
\vspace{10mm}
\plotone{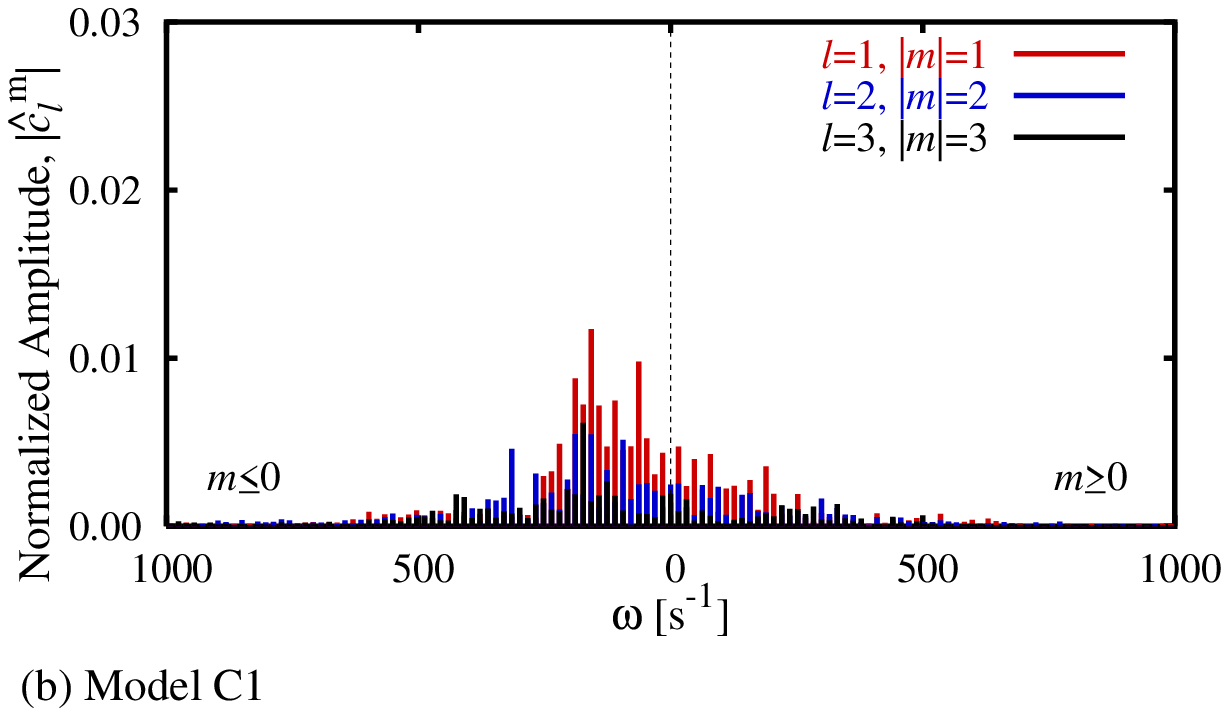}
\vspace{10mm}
\plotone{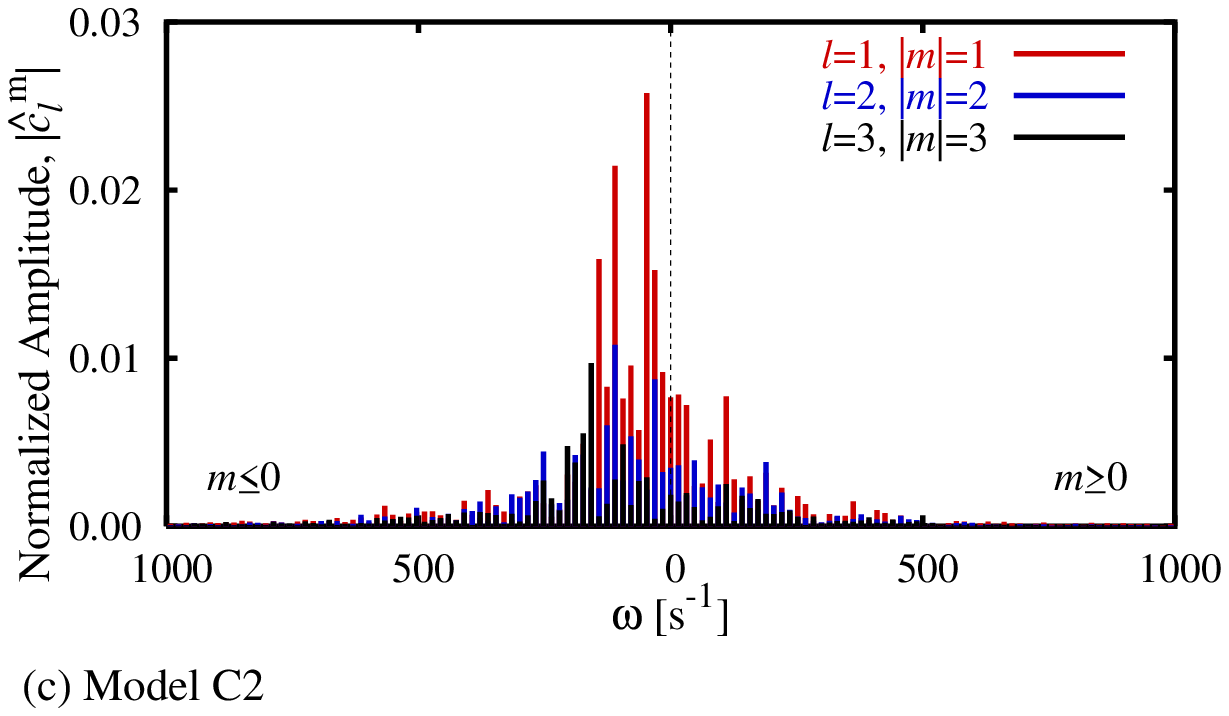}
\vspace{10mm}
\caption{
Fourier-transformed normalized amplitude $|\hat{c}^m_l(\omega)|$ of $|m|=l$ modes without rotation for (a) Model~C0 and with rotation for (b) Model~C1 and (c) Model~C2.
\label{sphfrandlm}}
\end{figure}

Figure~\ref{sphfrandlm} demonstrates the Fourier-transformed normalized amplitudes $|\hat{c}^m_l(\omega)|$ of $|m|=l$ modes for 3D Models C0, C1, and C2.
The basis function of $|m|=l$ modes did not have any node point in the $\theta$ direction.
The amplitudes of the $m\ge0$ modes are plotted on the right side of the graph, and those of the $m\le0$ modes are plotted on the left side. 
If the distribution of $|\hat{c}^m_l(\omega)|$ is completely symmetric with respect to $\omega = 0$ in this figure, this result can be interpreted in three ways: (1) only sloshing $\pm m$ modes (i.e., degenerate spiral $\pm m$ modes) grow; (2) a spiral $+m$ mode and a spiral $-m$ mode emerge with the same probability; (3) both (1) and (2) occur during sampling time.
On the other hand, an asymmetric distribution of $|\hat{c}^m_l(\omega)|$ suggests that the spiral $m$ modes with a larger amplitude is dominant.
Figure~\ref{sphfrandlm}~(a) plots the non-rotational 3D Model~C0,
revealing asymmetries in the distribution.
We recognized that spiral modes were generated in the randomly perturbed flow even for the non-rotational model.
However, the maximum amplitudes were roughly same between $m>0$ modes and $m<0$ modes.
We assumed that sloshing $\pm m$ modes dominated, or spiral $+m$ modes and spiral $-m$ modes would appear almost equally from 600ms to 1000 ms.
Figures~\ref{sphfrandlm}~(b) and (c) present the results of rotational Models C1 and C2.
The $m<0$ modes grew more than the $m>0$ modes.
Furthermore, the amplitude of the $m<0$ modes became larger with increasing rotation. 
We introduced the clockwise rotation in the same direction as $m<0$ modes.
Therefore, the faster the rotation, the larger the amplitudes of the mode rotating in the same direction as a globally rotational flow tends to be.

\section{Summary and Discussion}
We investigated the effects of rotation on SASI in the nonlinear phase with three-dimensional hydrodynamics simulations, for the purpose of application to the supernova core in the postbounce phase.
When the SASI entered the nonlinear phase, 
we imposed rigid rotation at the outer boundary of the iron core, 
whose specific angular momentum on the equatorial plane was assumed to agree with the recent stellar evolution calculations with magnetic fields.
After the rotational flow arrived at the shock wave,
rotation began to influence the SASI in the nonlinear phase.
Focusing on this stage,
we performed mode analysis for the nonspherical deformation of the shock front, using spherical harmonics in space and Fourier decompositions in time. 

First, we examined the effects of rotation on SASI for the axisymmetric flow in which only sloshing modes existed before rotation was added.
In the ranges of
rotational strength enough not to explode,
the average shock radius hardly changed with increasing angular momentum.
Moreover, mode analysis revealed that the sloshing modes were also insensitive to rotation except for the steady $l=2, m=0$ mode.
Combining our results with the outcomes obtained in the linear analyses by \citet{laming07} and \citet{yamasaki08},
rotation should barely affect the growth rates of sloshing modes.

Moreover, we studied the effects of rotation on SASI for the non-axisymmetric flow in which both sloshing and spiral modes existed before rotation was added.
In contrast with the axisymmetric models,
the shock radius tended to expand more with increased rotation.
As faster rotation was added to the flows,
spiral flows ran with higher velocity from the triple points to near the PNS, and then
flows circulated with higher velocity inside the high-entropy blobs. 
Large centrifugal force acted on the circulating flows inside the blobs, and the flows running in the rotational direction might push the shock wave further outward.
As with the axisymmetric models, 
the effect of rotation on the sloshing modes became prominent for the steady $l=2, m=0$ mode.
We observed that the other sloshing modes also grew slightly with increasing rotational velocity.
This result might be due to nonlinear effects of the coupling of the $m\neq0$ modes and $m=0$ modes.
For non-rotational models, sloshing $\pm m$ modes were dominant,
or spiral $+m$ modes and spiral $-m$ modes emerged with almost the same probability.
However, for rotational models, the spiral modes rotating in the same direction as the rotational flow developed significantly with faster rotation. 
These results agreed with the linear analyses of SASI by \citet{laming07} and \citet{yamasaki08}.

It should be noted that the simulations highlighted in this paper are a first step towards realistic three-dimensional modeling of supernova explosions.
The approximations adopted in this paper (e.g., the replacement of the PNS by the fixed inner boundary and the light-bulb approach with the constant neutrino luminosity) need improvement.
It is important to clarify how rotational SASI impacts neutrino heating,
but transport schemes beyond the light-bulb approaches are unquestionably needed.
Confirming the outcomes of this paper will require consistent simulations in 3D, covering the entire stellar core and starting from gravitational collapse with better neutrino transport,
which is computationally prohibitive at present.
Additionally, understanding the effects of rotation on the linear growth of SASI and constructing steady accretion flows with rotation are major undertakings.
The generation of pulsar spins by SASI has been addressed by \citet{blondin07a}.
Currently, we are systematically investigating a possible correlation between the kick and spin of PNS,
and our results will be presented in a forthcoming paper,
along with the discussions of magnetic effect on SASI (Iwakami et al. in preparation).

\acknowledgements
WI expresses her sincere gratitude to K. Ueno, and M. Furudate
for their continuing encouragement and suggestions.
KK is grateful to K. Sato for continuous encouragement. 
Numerical computations were performed on the Altix3700Bx2 at the Institute of Fluid Science, Tohoku University, as well as on XT4 and the general common use computer system at the center for Computational Astrophysics, CfCA, the National Astronomical Observatory of Japan.
This study was supported in part by JSPS Research Fellowships and was partially supported 
by the Program for Improvement of Research Environment for Young Researchers from Special Coordination Funds for Promoting Science and
Technology (SCF), the Grants-in-Aid for the Scientific Research (No. S19104006, No. S14102004, No. 14079202, No. 14740166, No. 20740150) and Grant-in-Aid for the 21st century COE program``Holistic Research and Education Center for Physics of Self-organizing Systems'' of Waseda University by the Ministry of Education, Culture, Sports, Science, and Technology (MEXT) of Japan.

\clearpage



\begin{table}
\begin{center}
\caption{Summary of all models.}
\vspace{4mm}
\begin{tabular}{clcc}
\tableline
\tableline
  Model & Perturbation & $\beta_\phi^a$ &
  $L$ [cm$^2$ s$^{-1}$]$^b$ \\
\tableline
 A0  & none       &  -     &  -                 \\
 A1  & none       & 0.0100 & $4.6\times10^{15}$ \\
 A2  & none       & 0.0125 & $5.8\times10^{15}$ \\
 A3  & none       & 0.0150 & $6.9\times10^{15}$ \\
 A4  & none       & 0.0200 & $9.2\times10^{15}$ \\
 B0  & $l$=1,$m$=0    &  -     &  -                 \\
 B1  & $l$=1,$m$=0    & 0.0100 & $4.6\times10^{15}$ \\
 B2  & $l$=1,$m$=0    & 0.0125 & $5.8\times10^{15}$ \\
 C0  & random     & -      &  -                 \\
 C1  & random     & 0.0100 & $4.6\times10^{15}$ \\
 C2  & random     & 0.0150 & $6.9\times10^{15}$ \\
\tableline
\end{tabular}
\tablenotetext{a}{Parameter for rotation in Eq. \ref{eq:rot}.}
\tablenotetext{b}{Specific angular momentum on the equatorial plane.}
\label{model}
\end{center}
\end{table}

\end{document}